\documentclass[runningheads]{llncs}
\usepackage[vietnamese, english]{babel}
\usepackage[T1]{fontenc}
\usepackage{graphicx}
\usepackage{booktabs} 
\usepackage{amsmath}
\usepackage{amssymb}

\usepackage{xcolor,listings}
\usepackage{textcomp}
\usepackage{color}
\newcommand{\VN}{\foreignlanguage{vietnamese}}
\usepackage[misc,geometry]{ifsym}

\begin{document}
\title{Improving Vietnamese Legal Document Retrieval using Synthetic Data}
%
%
\author{Pham Tien Son\inst{1} \and
Nguyen Doan Hieu\inst{1} \and Nguyen Dai An\inst{1} \and Dinh Viet Sang\inst{1}\thanks{Corresponding author}}

\authorrunning{P. T. Son et al.}
%

\institute{BKAI Research Center, School of Information and Communication Technology \\  Hanoi University of Science and Technology, Vietnam \\
\email{\{son.pt204891@sis, hieu.nd231135m, an.nd215296\}@sis.hust.edu.vn, sangdv@soict.hust.edu.vn}
}


%
\maketitle              

\begin{abstract}
In the field of legal information retrieval, effective embedding-based models are essential for accurate question-answering systems. However, the scarcity of large annotated datasets poses a significant challenge, particularly for Vietnamese legal texts. To address this issue, we propose a novel approach that leverages large language models to generate high-quality, diverse synthetic queries for Vietnamese legal passages. This synthetic data is then used to pre-train retrieval models, specifically bi-encoder and ColBERT, which are further fine-tuned using contrastive loss with mined hard negatives. Our experiments demonstrate that these enhancements lead to strong improvement in retrieval accuracy, validating the effectiveness of synthetic data and pre-training techniques in overcoming the limitations posed by the lack of large labeled datasets in the Vietnamese legal domain.

\end{abstract}
\section{Introduction}

The task of passage retrieval, which involves identifying relevant passages from a large corpus in response to a query, has become increasingly important with the rise of pre-trained language models like BERT \cite{devlin2018bert}. Enhancements such as Sentence-BERT \cite{reimers2019sentence} and SimCSE \cite{gao-etal-2021-simcse} have further improved text embedding techniques, leading to significant advances in Natural Language Processing (NLP).

For Vietnamese legal information retrieval, accurate and efficient retrieval systems are essential, particularly in legal question answering (QA) systems. Prior research has explored various models for this task: \cite{kien2020answering} introduced a new attention-based architecture, \cite{van2022miko} proposed combining BM25 with RoBERTa, and \cite{pham2022multi} used models like Sentence-BERT and coCondenser to enhance retrieval performance while \cite{pham2023question} utilized SimCSE and an ensemble reranker to produce state-of-the-art results on their benchmark data. Despite these efforts, a key limitation remains the scarcity of high-quality annotated datasets in the Vietnamese legal domain, hindering the development of robust retrieval systems.

To address this issue, we propose leveraging large language models (LLMs) to generate synthetic legal queries, creating a substantial dataset to improve passage retrieval. This synthetic data will be used to pre-train and fine-tune models like bi-encoder and ColBERT, enhancing their accuracy and scalability in the Vietnamese legal domain. Our approach not only addresses the immediate issue of data scarcity but also provides a scalable and efficient solution for improving legal information retrieval systems in the Vietnamese legal domain.

The primary contributions of this research are summarized as follows: 
\begin{enumerate}
    \item We present a method for generating synthetic queries based on passages of Vietnamese legal text, resulting in a dataset of 500,000 legal queries and corresponding passages;
    \item We implement the `Query-as-context Pre-training for Dense Passage Retrieval' \cite{Wu2022QueryascontextPF} technique for PhoBERT, further enhancing the retrieval performance of the backbone language model;
    \item We demonstrate improvements in retrieval accuracy through the application of bi-encoder and ColBERT retrieval models trained on the newly generated dataset.
\end{enumerate}

The rest of the paper is organized as follows: Section~\ref{sec:related} briefly summarizes prior works related to our work. Section~\ref{sec:propose} describes our method. In Section~\ref{sec:experiments}, we will present experimental results on benchmark datasets and compare them with the baseline and state-of-the-art methods. Section~\ref{sec:analyses} analyzes and discusses the results. Finally, we conclude the paper and discuss future work in Section~\ref{sec:conclude}.

\section{Related Work}

\label{sec:related}
\textbf{Text Embeddings.} Recently, there has been a shift of interest towards the use of neural retrieval techniques, which rely on dense vector representations to capture the underlying semantics of both queries and documents. Unlike traditional keyword-based approaches like BM25 or TF-IDF, neural models such as Sentence-BERT \cite{reimers2019sentence} and SimCSE \cite{gao-etal-2021-simcse} produce embeddings that enable semantic matching, allowing for more accurate retrieval based on context rather than exact term matching. These models leverage techniques like [CLS] token pooling or mean-pooling to aggregate embeddings, enabling fast and scalable comparisons via cosine similarity or dot product calculations. The contrastive loss is commonly used during training to fine-tune the models, aligning relevant query-document pairs while distancing irrelevant ones \cite{lin-etal-2023-train}.

Multi-vector models like ColBERT \cite{10.1145/3397271.3401075}, extend this approach by representing queries and documents with multiple vectors, providing more detailed interactions for improved retrieval accuracy. One notable development is the M3-Embedding model \cite{chen2024bge}, part of the BGE-M3 project, which supports dense, multi-vector, and sparse retrieval. This model’s versatility allows it to adapt to various retrieval scenarios, making it particularly beneficial for applications like legal text analysis. Additionally, M3-Embedding's multilingual capabilities, supporting over 100 languages, and its ability to handle input texts up to 8192 tokens, make it ideal for cross-lingual and large-scale document retrieval tasks.

Cross-encoders, on the other hand, enable even deeper query-document interaction by processing them together in a single transformer model, though their computational intensity limits their use to re-ranking stages in large-scale retrieval systems \cite{thakur2021beir}. Together, these methods represent the evolution of text embeddings in information retrieval, offering more precise and context-aware retrieval mechanisms than their lexical predecessors.

\subsubsection{Synthetic data.} One of the major obstacles to the widespread adoption of neural retrieval models is their requirement for large supervised training sets to surpass traditional term-based techniques, which are constructed from raw corpora \cite{ma-etal-2021-zero}. Thakur et al. \cite{thakur2021beir} find that BM25 remains a robust baseline for out-of-domain tasks. This highlights a critical challenge: in-domain performance cannot reliably predict how well an approach will generalize in a zero-shot setup. Many approaches that outperform BM25 on an in-domain evaluation perform poorly on the BEIR datasets.

Previous works have demonstrated that leveraging pretrained language models to generate queries can significantly enhance the performance of retrieval models in the absence of in-domain labeled data \cite{ma-etal-2021-zero,liang2020embedding,10.1145/3477495.3531863}. Motivated by a line of work on knowledge distillation from black-box LLMs through training on synthetic data generated from them, such as Orca \cite{mukherjee2023orca} and Phi \cite{gunasekar2023textbooks,wang2023improving} use GPT-3.5/4 \cite{achiam2023gpt} to generate query-document pairs across multiple tasks and languages. They then train open-source decoder-only LLMs on this synthetic data using standard contrastive loss, achieving state-of-the-art results in information retrieval benchmarks.

However, using large language models for information retrieval tasks as \cite{wang2023improving} is costly in practice, both in terms of computational resources and inference latency. Thus, this work examines the prospect of distilling the knowledge from these advanced LLMs into smaller language models such as BERT.

\subsubsection{Pre-training tailored for information retrieval.} Another approach to improve the performance of dense retrieval that has been studied is enhancing the pre-training process specifically for dense retrieval tasks. Techniques like Condenser \cite{gao2021condenser} involve adding a shallow decoder to the encoder, forcing it to reconstruct masked texts, thereby enhancing the encoder's capacity to produce meaningful embeddings. Based on this, coCondenser \cite{gao2021unsupervised} further improves performance by incorporating contrastive loss, which ensures that embeddings of similar text spans are brought closer together while those of different spans are pushed apart.

Another innovative approach is CoT-MAE \cite{10.1609/aaai.v37i4.25598}, which introduces a contextualized masked auto-encoder structure. This model uses both the target text and its context for reconstruction, further enhancing the encoder's understanding of the text within its broader context. Additionally, models like Query-as-Context \cite{Wu2022QueryascontextPF} extend these ideas by generating queries from passages for pre-training, which has shown promising results in improving retrieval performance. These advancements underscore the importance of specialized pre-training techniques in boosting the effectiveness of dense retrieval systems.

\section{Methodology}
\label{sec:propose}
\subsection{Overall}

\begin{figure}[htbp]
    \centering
    \includegraphics[width=\textwidth]{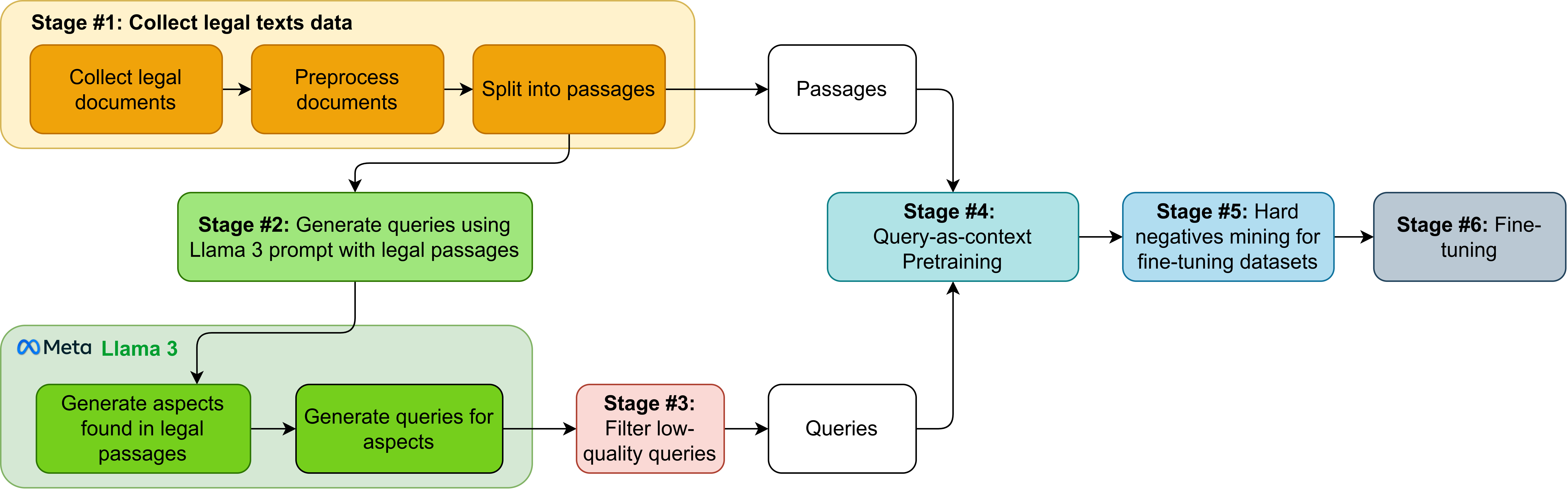}
    \caption{Workflow for generating synthetic queries and fine-tuning retrieval models using Vietnamese legal texts.} 
    \label{fig:overview}
\end{figure}

In this subsection, we outline our comprehensive workflow for generating synthetic query datasets and fine-tuning retrieval models using Vietnamese legal texts. This workflow is visually represented in Figure \ref{fig:overview}, which illustrates the sequential stages from data collection to model fine-tuning.

Our methodology is divided into several key stages:

\textbf{Stage 1: Collect legal text data.} This initial stage involves the collection and preprocessing of legal documents. The documents are split into smaller passages suitable for further processing.

\textbf{Stage 2: Generate queries using Llama 3 prompt with legal passages.} Using the Meta Llama 3 model, we generate synthetic queries based on aspects identified in the legal text passages. This involves generating aspects from the legal texts and then crafting queries for these aspects.

\textbf{Stage 3: Filter low-quality queries.} We remove low-quality queries that explicitly refer to the input passage and those that are only shallowly relevant to the input passage. For queries that are only shallowly relevant, we use the BGE-M3 dense retriever to filter out synthetic queries that cannot recover their input passage within the top 40 retrieved results.

\textbf{Stage 4: Query-as-Context Pre-training.} We employ the generated queries to further pre-train our language model, focusing on enhancing its ability to understand and retrieve relevant passages.

\textbf{Stage 5: Hard negatives mining for fine-tuning datasets.} We mine hard negative examples to create a robust fine-tuning dataset, which is crucial for improving the retrieval model's accuracy.

\textbf{Stage 6: Fine-tuning.} The final stage involves fine-tuning the pre-trained model with the generated data, optimizing it for the specific task of legal text retrieval.

\subsection{Data Curation}

Wang et al. \cite{wang2023improving} use GPT-3.5/4 to generate queries along with corresponding positive and hard negative passages by maintaining output diversity through two stages of generation: first generating retrieval tasks, then using them to generate query-passage pairs. As our objective focuses on creating a domain-specific dataset for Vietnamese legal text retrieval, relying solely on prompt engineering for this task would be complex and inefficient. Therefore, we opted to use collected legal text passages as input for an LLM to generate queries related to the content of each passage.

Our main source for collecting legal documents was thuvienphapluat.vn. The website hosts a wide range of legal documents, including Laws, Decrees, Circulars, Joint Circulars, Resolutions, Ordinances, Decisions, and the Constitution. After scraping both the metadata and full-text content, we preserved the hierarchical structure of each document, which typically includes chapters, sections, articles, and clauses. This structure was essential for maintaining context and ensuring that the text remained coherent after being split into smaller passages. Each passage retained crucial information, such as the document’s domain, title, header, and main content, which provided the necessary context for accurate query generation. This process resulted in 143,261 passages, which were used as input for the LLM to generate high-quality, contextually relevant queries.

\subsection{Synthetic Query Generation}
\label{sec:3_query_gen_process}

For generating synthetic queries, we chose the open-source LLM Llama 3 70B \cite{llama3modelcard} due to its strong performance, particularly in Vietnamese. Llama 3, trained on over 15 trillion tokens, outperforms many LLMs with a similar parameter count and demonstrates robust multilingual capabilities, making it well-suited for our needs.

\begin{figure}[t!]
    \centering
    \includegraphics[width=\textwidth]{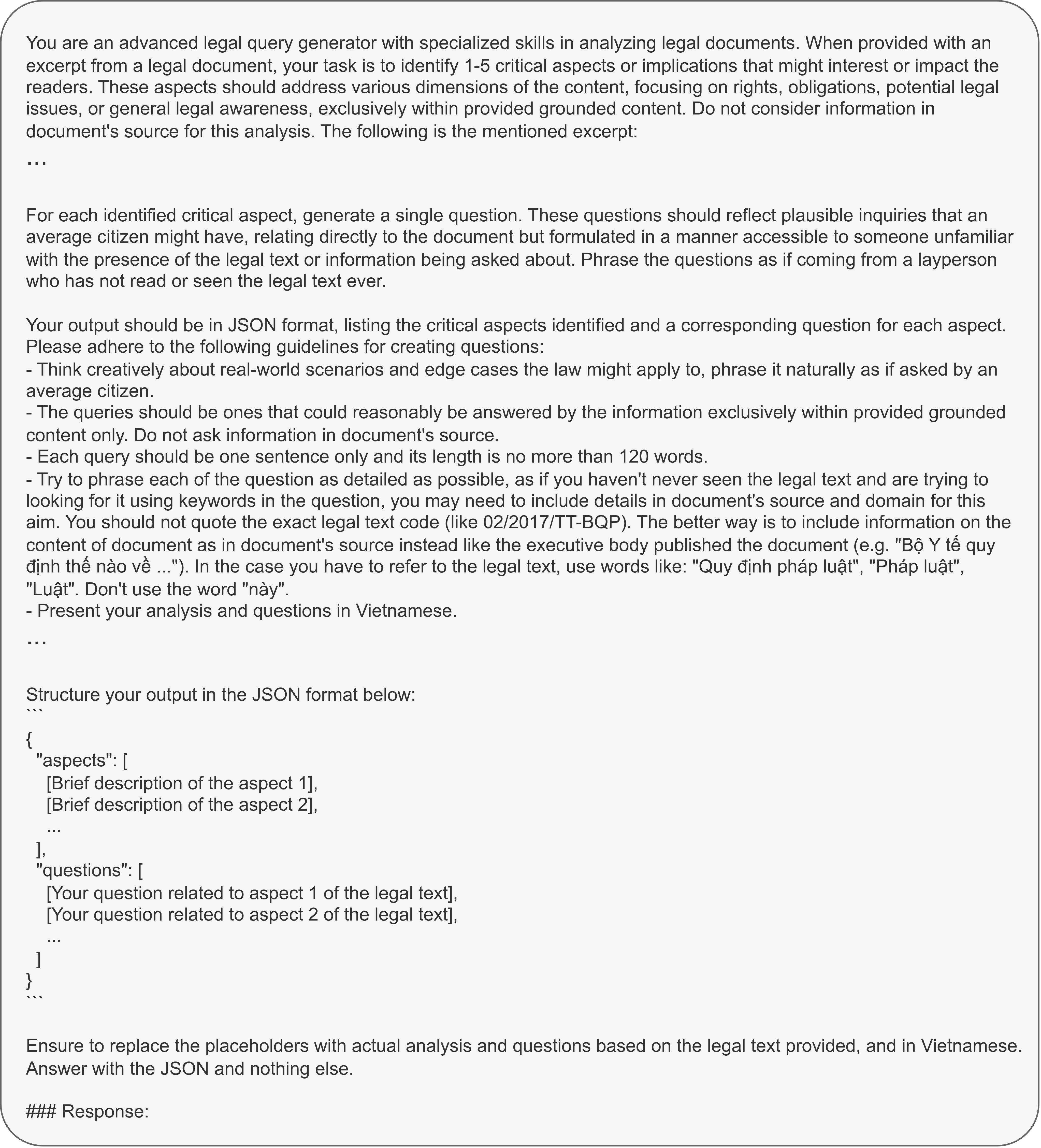}
    \caption{The shortened prompt template we used to generate synthetic queries from legal text passages, with placeholders for input documents and few-shot examples omitted.} 
    \label{fig:prompt_template}
\end{figure}

To maintain diversity and relevance in query generation, we experimented with different prompting techniques. A direct approach, where the model generated questions without identifying distinct aspects of the passage, often led to less diverse and sometimes irrelevant queries. Through various prompt designs, we discovered that instructing the model to identify 1 - 5 different aspects covered in the passage and then generate a question for each aspect yielded the most relevant and diverse queries. In Figure \ref{fig:prompt_template}, we show the prompt template used to generate synthetic queries from legal text passages. Examples of a generated query and its corresponding passage are shown in Table \ref{tab:example-query-passage}. A quantitative analysis of the generation method will be later discussed in Section \ref{sec:5.1}.

Applying this method, we generated over 620,000 legal queries from 140,292 passages extracted from our curated Vietnamese legal text collection. We then employed the BGE-M3 dense retriever \cite{chen2024bge}, which demonstrated strong zero-shot performance in our testing, to filter out queries whose corresponding passages did not appear in the top 40 relevant results. Additionally, we excluded queries that directly referred to the passage using terms like \VN{"quy định này"} or \VN{"thông tư này"}. This process results in a final dataset of 507,152 Vietnamese legal queries, covering a wide range of legal domains, which is then used to pre-train and fine-tune our retrievers.

We generated these synthetic queries using Llama 3 70B through Together AI's free credits program, with the entire generation process costing approximately 200 USD.

\begin{table}
    \centering
    \caption{Example of a generated query-passage pair for the domain \VN{"Tiền tệ - Ngân hàng"}.}
    \label{tab:example-query-passage}
    \resizebox{\textwidth}{!}{
    \begin{tabular}{l@{\hskip 0.1in}p{14cm}}
    \toprule
    \textbf{Domain} & \VN{Tiền tệ - Ngân hàng} \\
    \midrule
    \textbf{Header} & \VN{Mục 1. CHUẨN BỊ THANH TRA, Chương II. TRÌNH TỰ, THỦ TỤC TIẾN HÀNH CUỘC THANH TRA THEO KẾ HOẠCH THANH TRA, Thông tư 36/2016/TT-NHNN quy định về trình tự, thủ tục thanh tra chuyên ngành Ngân hàng do Thống đốc Ngân hàng Nhà nước Việt Nam ban hành.} \\
    \midrule
    \textbf{Content} & \VN{5. Trưởng đoàn thanh tra tổ chức họp Đoàn thanh tra để phổ biến kế hoạch tiến hành thanh tra được duyệt và phân công nhiệm vụ cho các Tổ thanh tra, Nhóm thanh tra, các thành viên của Đoàn thanh tra; thảo luận, quyết định về phương pháp, cách thức tổ chức tiến hành thanh tra; sự phối hợp giữa các thành viên Đoàn thanh tra, các cơ quan, đơn vị có liên quan trong quá trình triển khai thanh tra. Trong trường hợp cần thiết, người ra quyết định thanh tra hoặc người được người ra quyết định thanh tra ủy quyền dự họp và quán triệt mục đích, yêu cầu, nội dung thanh tra và nhiệm vụ của Đoàn thanh tra. Việc phân công nhiệm vụ cho các Tổ thanh tra, Nhóm thanh tra, các thành viên Đoàn thanh tra phải thể hiện bằng văn bản. \newline
    6. Tổ trưởng thanh tra, Nhóm trưởng thanh tra, thành viên Đoàn thanh tra xây dựng kế hoạch thực hiện nhiệm vụ được phân công và báo cáo Trưởng đoàn thanh tra trước khi thực hiện thanh tra tại tổ chức tín dụng.} \\
    \midrule
    \textbf{Aspect 1} & \VN{Trách nhiệm của Trưởng đoàn thanh tra trong việc tổ chức và phân công nhiệm vụ} \\
    \midrule
    \textbf{Query 1} & \VN{Ngân hàng Nhà nước quy định Trưởng đoàn thanh tra phải làm gì để chuẩn bị cho cuộc thanh tra?} \\
    \midrule
    \textbf{Aspect 2} & \VN{Quy trình xây dựng và báo cáo kế hoạch thực hiện nhiệm vụ của các Tổ thanh tra, Nhóm thanh tra} \\
    \midrule
    \textbf{Query 2} & \VN{Khi được phân công nhiệm vụ, các Tổ thanh tra, Nhóm thanh tra phải làm gì để chuẩn bị cho cuộc thanh tra?} \\
    \bottomrule
    \end{tabular}
    }
\end{table}

\subsection{Pre-training}

Query-as-Context pre-training \cite{Wu2022QueryascontextPF} is based on the observation that text spans within the same document can vary significantly in semantics, potentially weakening the effectiveness of traditional pre-training techniques. However, one limitation noted in this approach is that the T5 model used often produced a substantial number of unrelated query-passage pairs, which could diminish its overall effectiveness. This aligns well with the first part of our work — generating a legal query dataset — where we address this issue by utilizing an advanced LLM to produce queries and implementing a filtering process to remove irrelevant pairs, as detailed in the previous subsection.

Using the pairs of collected passages \( x_i \) and corresponding legal queries \( y_i \) generated by Llama 3, we apply the loss function from the Contextualized Masked Autoencoder (CoT-MAE) framework \cite{10.1609/aaai.v37i4.25598}. Specifically, the encoder reconstructs the passage \( x_i \) using its unmasked tokens, while the decoder reconstructs the query \( y_i \) by leveraging both its unmasked tokens and the contextual passage \( x_i \). The training objective combines the encoder’s masked language modeling (MLM) loss and the decoder’s context-supervised MLM loss, ensuring that the model learns to effectively integrate the query and passage context during pre-training.

\subsection{Fine-tuning}

To verify the effectiveness of our pre-training, we fine-tuned a bi-encoder and ColBERT model on downstream retrieval tasks. Our fine-tuning process is based on a single-stage pipeline with hard negative mining, utilizing a comprehensive set of datasets.

For both models, we utilized the MS-MARCO passage ranking dataset, SQuAD 2.0, 80\% of the Legal Text Retrieval Zalo 2021 training challenge dataset, and the dataset collected from thuvienphapluat.vn. Additionally, our synthetic query data generated for pre-training is also used in this fine-tuning process. For all datasets, we create hard negative passages for each training query from the BGE-M3 dense retrieval model. 

For each query \( q^+ \), the positive passage \( p^+ \) forms a pair \((h_{p^+}, h_{q^+})\). The negative samples \( \{p^-\} \) included hard negatives identified by the BGE-M3 dense retrieval and in-batch negative passages. The training objective for both models is to maximize the similarity between the query and the positive passage while minimizing the similarity between the query and the negative passages. This is achieved using the InfoNCE loss function:

\[
\mathcal{L} = - \log \frac{\exp(\text{sim}(h_{q^+}, h_{p^+}) / \tau)}{\sum_{p \in \{p^+, p^-\}} \exp(\text{sim}(h_{q}, h_{p}) / \tau)},
\] where \( \tau \) is a temperature hyper-parameter fixed to 1, and \( \text{sim}(\cdot, \cdot) \) represents the dot product similarity function.

\section{Experiment}
\label{sec:experiments}
\subsection{Datasets}

\begin{table}[tp]
    \centering
    \caption{Fine-tuning dataset statistics.}
    \label{tab:finetune-dataset-statistics}
    \begin{tabular}{lc@{\hskip 0.1in}r@{\hskip 0.1in}r@{\hskip 0.1in}r}
    \toprule
    \textbf{Training Dataset} & \textbf{Translated} & \textbf{\#Passage} & \textbf{\#Query} & \textbf{\#Positive Pairs} \\
    \midrule
    Zalo Legal 2021 (80\%) &   & 61,425   & 2,556   & 2,556   \\ 
    MS-MARCO \cite{nguyen2016ms} & \checkmark & 8,841,823 & 502,939 & 532,751 \\
    SQuAD 2.0 \cite{rajpurkar-etal-2018-know} & \checkmark & 13,317   & 60,942  & 60,942  \\ 
    TVPL & & 224,006 & 165,334 & 189,641 \\ 
    \bottomrule
    \end{tabular}
\end{table}

Table \ref{tab:finetune-dataset-statistics} presents the datasets used for fine-tuning and evaluation. Our primary fine-tuning datasets include MS-MARCO \cite{nguyen2016ms}, SQuAD 2.0 \cite{rajpurkar-etal-2018-know}, 80\% of the training set from the Legal Text Retrieval Zalo 2021 challenge (Legal Zalo 21), and the newly introduced TVPL dataset. Since MS-MARCO and SQuAD 2.0 are originally in English, we translated them into Vietnamese using Google Translate, following the approach of \cite{2024arXiv240301616Q}. The use of large, translated datasets has proven beneficial for improving the performance of monolingual retriever models due to the absence of comparably large annotated datasets for Vietnamese.

We developed TVPL — a new benchmark dataset for Vietnamese legal text retrieval. TVPL is named after the thuvienphapluat.vn website, from which its legal QA articles were sourced. The dataset comprises a training set of 165,334 queries and a test set of 10,000 queries, along with a corpus of 224,006 legal passages. This dataset addresses previous limitations of scale and diversity, providing a more comprehensive benchmark for evaluating retrieval models in the Vietnamese legal domain.

\begin{figure}[htbp]
    \centering
    \includegraphics[width=\textwidth]{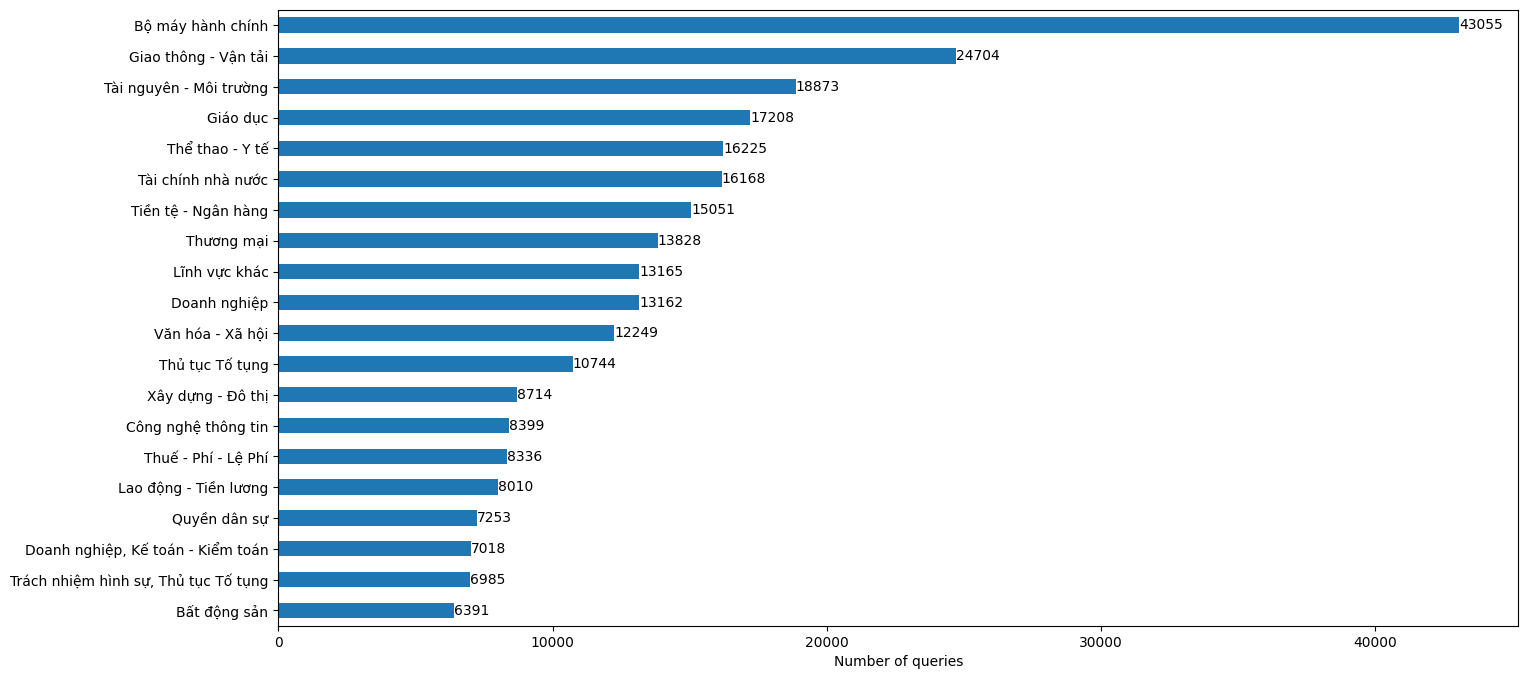}
    \caption{Top 20 domains by number of queries in synthetic query dataset.} 
    \label{fig:top20domain}
\end{figure}

Additionally, our 507,152 generated queries using the Llama3-70B model based on passages extracted from Vietnamese legal texts are also used in this fine-tuning stage. This synthetic dataset spans a broad range of legal domains, as shown in Figure \ref{fig:top20domain}.

For evaluation, we used 20\% of the Legal Zalo 21 dataset and 10,000 test queries from TVPL for in-domain testing. Additionally, we employed the Vietnamese Wiki Question Answering dataset from the Zalo AI Challenge 2019 for out-of-domain evaluation.

\subsection{Model Pre-training and Fine-tuning}

We used the synthetic dataset for pre-training, with text segmented by the Underthesea library. During pre-training, we select a batch of passages at each step and randomly choose a candidate query as context for each passage to form a relevant pair. The encoder for CoT-MAE was initialized with a pre-trained PhoBERT-base-v2 model, while the decoder was trained from scratch. Pre-training was conducted for 2,000 steps using the AdamW optimizer with a learning rate of 1e-4, a batch size of 1024, and a linear schedule on a TPU v4-8. After training, the decoder was discarded, and only the encoder was retained for fine-tuning.

For fine-tuning, we utilized the four annotated datasets presented in the previous subsections along with our generated synthetic data. Legal Zalo 21 passages were chunked to PhoBERT's 256-token limit. Hard negatives for each query were mined using the BGE-M3 dense retrieval model. The bi-encoder model was trained with a batch size of 64 across 4 TPU chips, optimized with AdamW at a 2e-5 learning rate for 170,000 steps (5 epochs), pairing each query with one positive passage and 7 hard negatives. ColBERT was trained with a batch size of 16, 15 hard negatives, and a longer training period of 290,000 steps (9 epochs), as the model continued improving throughout. ColBERT embeddings were compressed to 2 bits per dimension for evaluations.

\subsection{Baselines}
For evaluation, our baseline methods include both sparse and dense retrieval approaches, as outlined in Table \ref{tab:retrieval_performance}. The sparse retrieval baseline is represented by BM25. For dense retrieval baselines, we include results from the monolingual models vietnamese-sbert \cite{vietnamesesbert}  and vietnamese-bi-encoder \cite{2024arXiv240301616Q}, as well as the multilingual models BGE-M3 \cite{chen2024bge} and $\text{mE5}_{\text{base}}$ \cite{wang2024multilingual}.

\begin{table}[tp]
    \centering
\caption{Performance comparison on our TVPL and Legal Zalo 21 benchmarks.}
\label{tab:retrieval_performance}
\resizebox{\textwidth}{!}{%
\begin{tabular}{l|c@{\hskip 0.05in}c@{\hskip 0.1in}c@{\hskip 0.1in}c|c@{\hskip 0.05in}c@{\hskip 0.1in}c@{\hskip 0.1in}c}
\toprule
\textbf{Model} & \multicolumn{4}{c|}{\textbf{TVPL}} & \multicolumn{4}{c}{\textbf{Legal Zalo 21}} \\
 & MRR@10 & MAP@10 & R@10 & R@100 & MRR@10 & MAP@10 & R@10 & R@100 \\
\midrule
\textit{Sparse retrieval} & & & & & & & & \\
BM25 & 21.60 & 20.87 & 41.11 & 70.64 & 51.53 & 31.68 & 48.43 & 72.33 \\
\midrule
\textit{Dense retrieval} & & & & & & & & \\
vietnamese-sbert \cite{vietnamesesbert} & 48.93 & 45.89 & 74.37 & 92.87 & 46.31 & 32.74 & 48.95 & 71.88 \\
vietnamese-bi-encoder \cite{2024arXiv240301616Q} & 48.38 & 46.42 & 68.92 & 86.44 & 80.69 & 56.65 & 66.93 & 80.11 \\
$\text{mE5}_{\text{base}}$ \cite{wang2024multilingual} & 19.39 & 18.43 & 33.50 & 58.69 & 59.72 & 54.25 & \textbf{71.77} & \textbf{84.05} \\
BGE-M3 \cite{chen2024bge} & 32.68 & 31.32 & 51.78 & 74.46 & 64.43 & 44.02 & 57.28 & 75.28 \\
\midrule
\midrule
Bi-encoder & 70.37 & 67.96 & 87.09 & 96.34 & 79.31 & 57.90 & 68.02 & 81.76 \\
CoT-MAE Bi-encoder & \textbf{70.69} & \textbf{68.25} & \textbf{87.34} & \textbf{96.92} & \textbf{80.03} & \textbf{58.41} & \textbf{69.08} & \textbf{81.61} \\
\midrule
ColBERT & 73.90 & 71.39 & 88.68 & 96.46 & 84.15 & 60.54 & 67.93 & 80.84 \\
CoT-MAE ColBERT & \textbf{74.61} & \textbf{72.04} & \textbf{89.29} & \textbf{96.41} & \textbf{84.08} & \textbf{60.76} & \textbf{69.17} & \textbf{81.34} \\
\bottomrule
\end{tabular}
}
\end{table}

\subsection{Main Results}
As shown in Table \ref{tab:retrieval_performance}, the results demonstrate that fine-tuning with additional generated data improves retrieval performance across all evaluation metrics. Pre-training further enhances these results, with notable improvements in both the bi-encoder and ColBERT models. ColBERT, with its more granular multi-vector retrieval, achieves the highest scores across both the TVPL and Legal Zalo 21 benchmarks.

\begin{table}[tp]
    \centering
\caption{Out-of-domain evaluation on Vietnamese Wiki Question Answering dataset from the Zalo AI Challenge 2019.}
\label{tab:out_of_domain_performance}
\begin{tabular}{l|c@{\hskip 0.05in}c@{\hskip 0.1in}c@{\hskip 0.1in}c}
\toprule
\textbf{Model} & \multicolumn{4}{c}{\textbf{Zalo QA 19}}  \\
 & MRR@10 & MAP@10 & R@10 & R@100 \\
\midrule
\textit{Sparse retrieval} & & & & \\
BM25 & 45.33 & 42.54 & 67.56 & 87.17  \\
\midrule
\textit{Dense retrieval} & & & & \\
vietnamese-sbert \cite{vietnamesesbert} & 48.93 & 45.89 & 74.37 & 92.87 \\
vietnamese-bi-encoder \cite{2024arXiv240301616Q} & 68.15 & 64.81 & 85.39 & 96.20  \\
$\text{mE5}_{\text{base}}$ \cite{wang2024multilingual} & \textbf{72,76} & \textbf{70.03} & \textbf{91.55} & \textbf{98.57}  \\
BGE-M3 \cite{chen2024bge} & \textbf{76.69} & \textbf{74.37} & \textbf{94.26} & \textbf{99.06}  \\
\midrule
\midrule
Bi-encoder & 69.57 & 66.36 & 86.60 & 96.91   \\
CoT-MAE Bi-encoder & 68.22 & 64.91 & 85.38 & 96.20  \\
\midrule
ColBERT &  70.34 & 67.58 & 88.72 & 96.86 \\
CoT-MAE ColBERT & \textbf{72.38} & \textbf{69.72} & \textbf{90.47} & \textbf{97.71}  \\
\bottomrule
\end{tabular}
\end{table}

\subsection{Out-of-domain Evaluation}
To further evaluate the robustness and generalization capabilities of our models, we conducted an out-of-domain evaluation on the Vietnamese Wiki Question Answering dataset from the Zalo AI Challenge 2019. This dataset contains query-passage pairs covering a wide range of topics beyond the legal domain, on which our models were specifically pre-trained and fine-tuned.

Despite being trained solely in the legal context, our models demonstrated improved performance on the out-of-domain dataset, as shown in Table \ref{tab:out_of_domain_performance}. Notably, the pre-trained and fine-tuned ColBERT model achieves scores close to those of larger multilingual retrievers, such as $\text{mE5}_{\text{base}}$ and BGE-M3. These models, with significantly larger parameter counts ($\text{mE5}_{\text{base}}$ at 270M and BGE-M3 at 560M, compared to our models with 124M), are still strong candidates for zero-shot retrieval tasks. The results suggest that our approach, though specialized for legal text, maintains strong generalization capabilities in broader retrieval scenarios.

\section{Analyses}
\label{sec:analyses}
\subsection{Effects of Aspect-guided Query Generation Prompt}
\label{sec:5.1}

We analyze the improvements brought by the aspect-guided query generation method (Section \ref{sec:3_query_gen_process}) compared to basic prompting, where the LLM generates queries directly from input passages. Performance is evaluated using passage hit rate (the percentage of queries retrieving their corresponding passage) and document hit rate (the percentage of queries retrieving the correct document). We use the BGE-M3 dense retriever \cite{chen2024bge} to rank the top-$k$ relevant passages for each query.

As shown in Table \ref{tab:top_k_prompts}, the aspect-guided method significantly outperforms basic prompting across different top-$k$ values (10, 20, 40). At $k=10$, it achieves an 82.06\% passage hit rate and 91.60\% document hit rate, compared to 8.26\% and 87.23\%, respectively, for basic prompting.

\begin{table}[tp]
    \centering
\caption{Passage hit rate and document hit rate for different top-$k$ values of 10.000 queries generated by Llama 3 using two prompting methods.}
\label{tab:top_k_prompts}
\resizebox{\textwidth}{!}{%
\begin{tabular}{lcccccccc}
\toprule
\textbf{Prompt} & \multicolumn{2}{c}{$k=10$} & \multicolumn{2}{c}{$k=20$}  & \multicolumn{2}{c}{$k=40$} \\
 & Passage hit & Document hit & Passage hit & Document hit & Passage hit & Document hit \\
 \midrule
Basic prompt & 8.26 & 87.23 & 11.93 & 91.05 & 16.15 & 93.38 \\
Aspect-guided query& \textbf{82.06} & \textbf{91.60} & \textbf{87.92} & \textbf{94.51} & \textbf{91.90} & \textbf{96.30}  \\
 generation prompt & \\
\bottomrule
\end{tabular}
}
\end{table}

\subsection{Quality - Space Footprint Trade-off: ColBERT's Residual Compression}

We examine the impact of increasing the number of bits for compression on both retrieval accuracy and storage requirements. Experiments were conducted on 224,006 passages from the TVPL dataset, with evaluation metrics on its test set. ColBERT's residual compression, as proposed in \cite{santhanam2021colbertv2}, offers improvements in both storage and retrieval performance. While higher bit sizes improve performance slightly, they come with a significant increase in storage. In contrast, The 1-bit ColBERT configuration performs competitively with minimal storage (647.33MB), even outperforming bi-encoder (672.02MB) in metrics like MRR@10 and MAP@10.

\begin{table}[tp]
\centering
\caption{Impact of compression on storage and retrieval performance on TVPL benchmark.}
\label{tab:compress}
\resizebox{\textwidth}{!}{
    \begin{tabular}{l@{\hskip 0.1in}c@{\hskip 0.1in}c@{\hskip 0.1in}c@{\hskip 0.1in}c@{\hskip 0.1in}c@{\hskip 0.1in}c@{\hskip 0.1in}c@{\hskip 0.1in}c}
    \toprule
    Index & Storage (MB) & F2@10 & MRR@10 & MAP@10 & Recall@10 & Recall@100 &  \\
    \midrule
    \textit{ColBERT} \\
    1 bit & 647.33 & 0.3371 & 73.61 & 71.04 & 88.49 & 96.24  \\
    2 bits & 1094.04 & 0.3406 & 74.61 & 72.04 & 89.29 & 96.41  \\
    4 bits & 1987.48 & 0.3411 & 74.93 & 72.34 & 89.43 & 96.51  \\
    8 bits & 3774.34 & 0.3407 & 75.02 & 72.43 & 89.31 & 96.49  \\
    \midrule
    \midrule
    \textit{Bi-encoder} & 672.02 & 0.3407 & 70.69 & 68.25 & 87.34 & 96.92 \\
    \bottomrule
    \end{tabular}
}
\end{table}

\section{Conclusion and Future Work}
\label{sec:conclude}

In this work, we proposed methods to improve Vietnamese legal text retrieval using synthetic data. Our key contributions include generating synthetic legal queries from Vietnamese legal text passages with a pre-trained LLM, creating a dataset of 500K query-passage pairs, and significantly enhancing retrieval accuracy with bi-encoder and ColBERT models trained on this dataset. Our experiments demonstrate the effectiveness of fine-tuning with synthetic data for improving model performance. We also applied the query-as-context CoT-MAE pre-training technique, which further boosted retrieval accuracy. The combination of synthetic data and CoT-MAE pre-training consistently yielded superior performance in both in-domain and out-of-domain evaluations.

The dataset generated in this work has been made publicly available on Hugging Face under the CC BY 4.0 license\footnote{\url{https://huggingface.co/datasets/phamson02/large-vi-legal-queries}}. Additionally, we are also publishing the TVPL benchmark dataset queries\footnote{\url{https://huggingface.co/datasets/phamson02/tuvanphapluat}}. We hope that these datasets and the method introduced can help advance the development of new language models for Vietnamese, potentially extending beyond legal contexts, and support applications like legal search and question-answering systems to benefit both public servants and citizens.

In future work, we want to explore the potential of using generated queries as inputs for large language models to create an artificial legal corpus. This corpus could, in turn, produce additional queries, creating a cycle to expand the dataset further. A deeper qualitative analysis comparing synthetic data to real-world data is necessary to understand this approach's strengths and limitations better. We also aim to examine potential performance collapse in models primarily trained on synthetic data and investigate strategies to mitigate this risk.

\begin{credits}
\subsubsection{\ackname} This work was funded by the NAVER Corporation within the framework of collaboration with the International Research Center for Artificial Intelligence (BKAI), School of Information and Communication Technology, Hanoi University of Science and Technology. We also extend our gratitude to Google’s TPU Research Cloud program for providing the necessary resources to run the experiments. This work would not have been possible without their contributions. Nguyen Doan Hieu was funded by the Master, PhD Scholarship Program of Vingroup Innovation Foundation (VINIF), code VINIF.2022.ThS.BK.09.

\end{credits}
%
%
%
\bibliographystyle{splncs04}
\bibliography{reference}
%




\end{document}